\documentclass[aps,prd,preprintnumbers,nofootinbib,floatfix,11pt]{revtex4}

\usepackage{amsfonts,amsmath,amssymb,mathrsfs,graphicx,epsfig,color,amsthm} 
\usepackage{hyperref}

\allowdisplaybreaks[1]
\newcommand{\be}{\begin{equation}}
\newcommand{\ee}{\end{equation}}
\newcommand{\ba}{\begin{eqnarray}}
\newcommand{\ea}{\end{eqnarray}}

\theoremstyle{definition}
\newtheorem{theorem}{Theorem}

\newtheorem{definition}[theorem]{Definition}

\usepackage{}

\usepackage[normalem]{ulem}

\begin{document}

\title{Four types of attractive gravity probe surfaces}
\author{Kangjae Lee${}^1$, Tetsuya Shiromizu$^{1,2}$, Keisuke Izumi$^{2,1}$, Hirotaka Yoshino$^3$ and Yoshimune Tomikawa$^4$}

\affiliation{$^{1}$Department of Mathematics, Nagoya University, Nagoya 464-8602, Japan}
\affiliation{$^{2}$Kobayashi-Maskawa Institute, Nagoya University, Nagoya 464-8602, Japan} 
\affiliation{$^{3}$Department of Physics, Osaka Metropolitan University, Osaka 558-8585, Japan}
%\affiliation{$^{4}$Department of Physics, Kobe University, Kobe 657-8501, Japan}
\affiliation{$^{4}$Division of Science, School of Science and Engineering, Tokyo Denki University, Saitama 350-0394, Japan}

\begin{abstract}
\begin{center}
{\bf Abstract}
\end{center}
\noindent
We reexamine the concept of the attractive gravity probe surface recently proposed as an indicator for strength of gravity. Then, 
we propose three new variant concepts and show refined inequalities
for the four types of the AGPSs by taking account of the angular 
momentum, gravitational waves and matters. 
\end{abstract}

\maketitle

%%%%%%%%%%%%%%%%%%%%%%%%%%%%%%%%%%%%%%%%%%%%%%%%%%%%%%%%%%%%%%%%%%%%%%%%%%%%%%%%%%%%%%%%%%%%%%%%%
%%%%%%%%%%%%%%%%%%%%%%%%%%%%%%%%%%%%%%%%%%%%%%%%%%%%%%%%%%%%%%%%%%%%%%%%%%%%%%%%%%%%%%%%%%%%%%%%%

\section{Introduction}

For a spacetime including black holes, Penrose proposed an inequality  which 
gives the upper bound for the black hole size as \cite{Penrose:1973um}
\begin{equation}
A_{\rm H}\le4\pi\left(2m\right)^2,
\end{equation} 
where $A_{\rm H}$ is the area of the event horizon and
$m$ is the Arnowitt-Deser-Misner (ADM) mass. 
This inequality has been proven for apparent horizons
in time-symmetric initial data \cite{wald, imcf, bray}. 
If one wants to include the contribution from angular momentum $J$, 
the Penrose inequality would be refined as 
\begin{equation}
\label{penroseineq}
A_{\rm H} \le A_{\rm Kerr}:=8\pi m \left(m+\sqrt{m^2-J^2/m^2}\right),
\end{equation}
where $A_{\rm Kerr}$ is the area of the event horizon of the Kerr black hole. Here we remind of the fact that the inequality 
$A_H \geq 8 \pi J $ has been proven under certain assumptions \cite{Dain2011a, Dain2011b, Dain2011c}. 
If one can show $A_H \geq 8\pi J$ for general cases, Eq.~\eqref{penroseineq} implies
\begin{equation}
m^2 \geq \Bigl( \frac{{\cal R}_H}{2} \Bigr)^2 +\frac{J^2}{{\cal R}_H^2}, \label{conj2}
\end{equation}
where ${\cal R}_H$ is the areal radius, ${\cal R}_H={\sqrt {A_H/4\pi}}$.
This inequality has been 
addressed under certain assumptions \cite{Anglada:2017ryp, Anglada2020, Khuri2018, Kopinski2020, Dain2018}. 

If the cosmic censorship conjecture holds, that is,
if any singularity is enclosed by an event horizon,
any spacetime with a horizon will 
settle down to a stationary Kerr black hole spacetime
due to the uniqueness theorem \cite{uniq}. Together with 
the area theorem \cite{hawking}, one can expect that Eq.~\eqref{conj2}
would hold. Although giving a direct proof of  
the cosmic censorship conjecture would be difficult,
the validity of 
the Penrose inequality gives us a strong support
for the cosmic censorship conjecture. 

The concepts of the horizons are important 
in studying the property of black hole physics theoretically. However, 
distant observers cannot see black holes by definition. 
Therefore, if one takes into account observations,
it is better to introduce an alternative geometrical concept
which describes strong gravity outside the horizon. Inspired by the feature of 
the Schwarzschild spacetime, the concepts of
the loosely trapped surface (LTS) \cite{Shiromizu:2017ego} and the dynamically 
transversely trapping surface (DTTS) \cite{Yoshino:2019dty} have been proposed.
These two concepts may be regarded as generalizations of the photon surface \cite{Claudel:2000yi}. 
While the former is formulated by the behavior  
of the trace of the extrinsic curvature of a two-surface in
a spacelike hypersurface\footnote{The trace of the extrinsic curvature is proportional to the expansion rate 
of null geodesic congruence on time-symmetric initial data.},
the latter is based on the behaviors of null geodesics tangent 
to the surface. 
The Penrose-like inequalities have been proven for these two surfaces
under certain assumptions. 
For certain cases, using the inverse mean curvature flow,
the inequalities for the areas of the LTS and DTTS, $A_{\rm LTS/DTTS}$, 
are given by
\begin{equation}
A_{\rm LTS/DTTS}\,\le\, 4\pi\left(3m\right)^2.
\end{equation}
For Einstein-Maxwell systems, the inequality
that includes the contributions from electromagnetic fields also has been
addressed \cite{Lee:2020pre}. 

As the extended concept of the LTS,
the attractive gravity probe surface (AGPS) 
was proposed in Ref.~\cite{Izumi:2021hlx}.
The AGPS characterizes the strength of gravity
on a surface 
by introducing an intensity parameter $\alpha$.
For a special case of $\alpha=0$, the AGPS is reduced to the LTS.
Therefore, this concept can be used to characterize
the strength of gravity for a wide range of spacetime: from
asymptotically flat region ($\alpha\to-1/2$) to the minimal surface ($\alpha\to\infty$).
It was shown that the AGPS also 
satisfies the areal inequality that includes the parameter $\alpha$~\cite{Izumi:2021hlx}.
Soon after the proposal of Ref.~\cite{Izumi:2021hlx},
the effect of the angular momentum, 
gravitational waves and matters to the areal inequalities
for the AGPS was examined \cite{Lee:2021hft}. 

The concept of the AGPS is defined by requiring  
the first derivative of the mean curvature 
in the normal direction divided by the squared mean curvature
to be equal to or larger than $\alpha$ [see Eq.~\eqref{def-lts}].
However, after the work of Ref.~\cite{Izumi:2021hlx},
we have noticed that there is an alternative way to generalize
the LTS.
Furthermore, these ideas of two extensions can also be applied to
the DTTS. 
In this paper, therefore, we propose three variants of the AGPS:
one is the extension of the LTS in a different method
(which we will call the longitudinal AGPS associated with the Ricci scalar, LAGPS-r),
and the other two are the extensions of the DTTS
in two ways (which we will call the transverse AGPS associated with the mean curvature, TAGPS-k, and the
transverse AGPS associated with the Ricci scalar, TAGPS-r).
The original AGPS proposed in \cite{Izumi:2021hlx}
is denoted as the longitudinal AGPS associated with the mean curvature
(LAGPS-k) in this paper. 
In total, there are four types of the AGPS 
and we examine their areal inequalities.

The rest of this paper is organized as follows. In Sec.~\ref{preparation}, we will describe our setting and present 
some geometrical formulas which are used in almost all of the proof for the Penrose-like inequalities. In Sec.~\ref{lts}, we will
review the concept of the original AGPS (i.e., LAGPS-k in this paper)
and give an  
alternative extension of the LTS (i.e., LAGPS-r). This part includes the previous studies \cite{Izumi:2021hlx, Lee:2021hft}. 
Next, we will present two extensions of the DTTS (i.e., TAGPS-k and TAGPS-r)
in Sec.~\ref{tagps}.
Section~\ref{summary} is devoted to a summary and discussions.
In Appendix A, we will present 
a certain unified treatment of the LAGPS-k and the LAGPS-r,
and their Penrose-like inequalities.
In Appendix B, the similar discussion is given for
the TAGPS-k and the TAGPS-r.
Throughout the paper, we use the unit in which
the speed of light and the gravitational constant
are unity, $c=G=1$.

%%%%%%%%%%%%%%%%%%%%%%%%%%%%%%%%%%%%%%%%%%%%%%%%%%%%%%%%%%%%%%%%%%%%%%%%%%%%%%%%%%%%%%%%%

\section{Preliminaries}
\label{preparation}

This section corresponds to the preparation. 
Firstly we will describe our setup and present some geometrical formula in subsection \ref{formula}. 
Then, we will review one of the key equations obtained through the monotonicity of Geroch's energy 
on the inverse mean curvature flow in subsection \ref{geroch-mono}. In subsection \ref{vacuum}, 
some basics of vacuum and axisymmetric spacetimes will be shown. 

\subsection{Setup and some geometrical formula}
\label{formula}

In a four-dimensional spacetime $(M,g_{ab})$, we consider a codimension-one spacelike hypersurface 
$\left(\Sigma,q_{ab},K_{ab}\right)$ and a codimension-one timelike hypersurface $\left(S,p_{ab},\bar{K}_{ab}\right)$ 
with the spacelike two-surface intersection $(\sigma_0, h_{ab})$ with $\Sigma$, 
where 
$K_{ab}$ and $\bar{K}_{ab}$ are the extrinsic curvatures of $\Sigma$ and $S$, respectively. 
For our current purpose, we
restrict the situation and suppose $\Sigma$ to be orthogonal to $S$. 
With the future directed timelike unit normal vector $n^a$ to $\Sigma$ and  the outward spacelike unit 
normal vector $r^a$ to $S$, the metric $g_{ab}$ is decomposed as 
\begin{equation}
g_{ab}=q_{ab}-n_an_b=p_{ab}+r_ar_b=h_{ab}-n_an_b+r_ar_b.
\end{equation}  
The extrinsic curvature of $\Sigma$ and $S$ are written by 
\begin{equation}
K_{ab}=\frac{1}{2}{\mbox \pounds}_nq_{ab}
\end{equation}
and
\begin{equation}
\bar{K}_{ab}=\frac{1}{2}{\mbox \pounds}_r p_{ab},
\end{equation}
respectively, where ${\mbox \pounds}$ is the
Lie derivative in $(M,g_{ab})$. For $(\sigma_0,h_{ab})$, one can 
define the two extrinsic curvatures as 
\begin{equation}
k_{ab}=\frac{1}{2}{}^{(3)}{\mbox \pounds}_{r}h_{ab}
\end{equation}
and
\begin{equation}
\kappa_{ab}=\frac{1}{2}{}^{(3)}\bar{\mbox \pounds}_{n}h_{ab},
\end{equation}
where ${}^{(3)}{\mbox \pounds}$ and ${}^{(3)}\bar{\mbox \pounds}$ 
are the Lie derivatives in $\Sigma$ and $S$ respectively. 

From the definition of the Ricci tensor, 
the Lie derivative of the trace of $\kappa_{ab}$, $\kappa$, is written as 
\begin{equation}
\label{liederiv.kappa1}
{}^{(3)}\bar{\mbox \pounds}_n\kappa=-{}^{(3)}\bar{R}_{ab}n^an^b-\kappa_{ab}\kappa^{ab}+N^{-1}{\cal D}^2N.
\end{equation}
Here, $N$ is the lapse function with respect to $n^a$, i.e. $n_a=-N (dt)_a$ where the coordinate $t$ is chosen so that a $t$-constant hypersurface
corresponds to $\Sigma$, 
${}^{(3)}\bar{R}_{ab}$ is the Ricci tensor of $(S,p_{ab})$ and ${\cal D}_a$ is the covariant derivative with respect to $(\sigma_0,h_{ab})$. 
Taking the double trace for the Gauss equation on $\sigma_0$ in $S$ and $S$ in $M$, we have 
\begin{equation}
{}^{(2)}R={}^{(3)}\bar{R}+2{}^{(3)}\bar{R}_{ab}n^an^b-\kappa^2+\kappa_{ab}\kappa^{ab}\label{gauss.eq1}
\end{equation}
and
\begin{equation}
{}^{(3)}\bar{R}=-2G_{ab}r^ar^b+\bar{K}^2-\bar{K}_{ab}\bar{K}^{ab},\label{gauss.eq1.2}
\end{equation}
where ${}^{(2)}R$, ${}^{(3)}\bar{R}$ and $G_{ab}$ are the Ricci scalar of $(\sigma_0,h_{ab})$, 
Ricci scalar of $(S,p_{ab})$ and the Einstein tensor for $(M,g_{ab})$, respectively. Then, 
with Eqs.~\eqref{gauss.eq1} and \eqref{gauss.eq1.2},  
Eq.~\eqref{liederiv.kappa1} is rewritten as 
\begin{equation}
\label{liederiv.kappa2}
{}^{(3)}\bar{\mbox \pounds}_n\kappa=-\frac{1}{2}{}^{(2)}R-G_{ab}r^ar^b+\frac{1}{2}\left(\bar{K}^2-\bar{K}_{ab}\bar{K}^{ab}
-\kappa^2-\kappa_{ab}\kappa^{ab}\right)+N^{-1}{\cal D}^2N.
\end{equation}

Similarly to Eq. (\ref{liederiv.kappa1}), the Lie derivative of the trace of $k_{ab}$, i.e. $k$, is written as
\begin{equation}
{}^{(3)}{\mbox \pounds}_rk=-{}^{(3)}R_{ab}r^ar^b-k_{ab}k^{ab}-\varphi^{-1}{\cal D}^2\varphi. \label{liederiv.k2}
\end{equation}
Here, $\varphi$ is the lapse function with respect to $r^a$, i.e. $r_a=\varphi (dy)_a$ where the coordinate $y$ is chosen so that
a $y$-constant hypersurface corresponds to $S$, and ${}^{(3)}R_{ab}$ is the 
Ricci tensor of $(\Sigma,q_{ab})$. 
With the double trace of the Gauss equation on $\sigma_0$ in $\Sigma$ and $\Sigma$ in $M$, 
\begin{equation}
{}^{(2)}R={}^{(3)}R+2{}^{(3)}\bar{R}_{ab}r^ar^b+k^2-k_{ab}k^{ab} \label{gauss.eq3}
\end{equation}
and
\begin{equation}
{}^{(3)}R=2G_{ab}n^an^b-K^2+K_{ab}K^{ab}, \label{gauss.eq2}
\end{equation}
Eq. (\ref{liederiv.k2}) becomes 
\begin{equation}
\label{liederiv.k}
{}^{(3)}{\mbox \pounds}_rk=\frac{1}{2}{}^{(2)}R-G_{ab}n^an^b+\frac{1}{2}\left(K^2-K_{ab}K^{ab}
-k^2-k_{ab}k^{ab}\right)-\varphi^{-1}{\cal D}^2\varphi.
\end{equation}

We decompose the extrinsic curvature $\bar{K}_{ab}$ and $K_{ab}$ as 
\begin{align}
\bar{K}_{ab}=\bar{K}_{(n)}n_an_b+k_{ab}+v_an_b+v_bn_a \label{dec.ext.curv1}
\end{align} 
and
\begin{align}
K_{ab}=K_{(r)}r_ar_b+\kappa_{ab}+v_ar_b+v_br_a,\label{dec.ext.curv2}
\end{align} 
where $\bar{K}_{(n)}:=\bar{K}_{ab}n^an^b$, $K_{(r)}:=K_{ab}r^ar^b$ 
and $v_a:=h_{ab}r^cK_c{}^b=-h_{ab}n^c\bar{K}_c{}^b$. Then, 
Eqs.~\eqref{liederiv.kappa2} and \eqref{liederiv.k} become 
\begin{align}
\label{liederiv.kappa3}
{}^{(3)}\bar{\mbox \pounds}_n\kappa&=-\frac{1}{2}{}^{(2)}R-G_{ab}r^ar^b-k\bar{K}_{(n)}+\frac{1}{2}\left(k^2-k_{ab}k^{ab}
-\kappa^2-\kappa_{ab}\kappa^{ab}\right)+v_av^a+N^{-1}{\cal D}^2N
\end{align} 
and
\begin{align}
\label{liederiv.k3}
{}^{(3)}{\mbox \pounds}_rk&=\frac{1}{2}{}^{(2)}R-G_{ab}n^an^b+\kappa K_{(r)}+\frac{1}{2}\left(\kappa^2-\kappa_{ab}\kappa^{ab}
-k^2-k_{ab}k^{ab}\right)-v_av^a-\varphi^{-1}{\cal D}^2\varphi.
\end{align} 
Note that we have not used the Einstein equation up to now,
and the above equations are purely geometric relations. 

%%%%%%%%%%%%%%%%%%%%%%%%%%%%%%%%%%%%%%%%%%%%%%%%%%%%%%%%%%%%%%%%%%%%%%%%%%%

\subsection{Geroch energy and inverse mean curvature flow}
\label{geroch-mono}

We assume that
it be possible to take a smooth inverse mean curvature flow (IMCF)
$\lbrace \sigma_y \rbrace_{y \in {\bf R}}$ on the 
spacelike hypersurface $\Sigma$, that is, $k \varphi=1$. In addition, 
we suppose that each foliation $\sigma_y$ is homeomorphic to a two-sphere. 
Then, assuming $\Sigma$ to be a maximal slice, 
one can show that the monotonicity of the Geroch energy defined by
\begin{equation}
\label{Geroch}
E(y):=\frac{A^{1/2}(y)}{64\pi^{3/2}}\int_{\sigma_y}\left(2{}^{(2)}R-k^2\right)dA,
\end{equation}
where $A(y)$ is the area of $\sigma_y$. 
In the IMCF, the first derivative of the Geroch energy with respect to $y$ is calculated as \cite{geroch}
\begin{equation}
\label{deriv.Geroch}
\frac{dE}{dy}=\frac{A^{1/2}}{64\pi^{3/2}}\int_{\sigma_y}\Bigl[ 2 \varphi^{-2}({\cal D} \varphi)^2
+{}^{(3)}R+\tilde k_{ab}\tilde k^{ab} \Bigr] dA,
\end{equation}
where $\tilde k_{ab}$ is the traceless part of $k_{ab}$,
i.e. $\tilde k_{ab}=k_{ab}-(k/2)h_{ab}$. 
Using Eqs.~\eqref{gauss.eq2} and \eqref{dec.ext.curv2} with
the Hamiltonian constraint
of the Einstein equations, $G_{ab}n^a n^b=8\pi T_{ab} n^a n^b$,
we rewrite the three-dimensional Ricci scalar ${}^{(3)}R$ of $\Sigma$ as 
\begin{equation}
{}^{(3)}R=16\pi\rho+2v_av^a+\tilde{\kappa}_{ab}\tilde{\kappa}^{ab}-2\kappa K_{(r)}-\frac{1}{2}\kappa^2, \label{3ricci}
\end{equation}
where $\tilde \kappa_{ab}$ is the traceless part of $\kappa_{ab}$ and 
$\rho$ is the energy density of matters, that is, $\rho:=T_{ab}n^an^b$.
%We now assume  
%$\Sigma$ to be a maximal slice (i.e., $K=K_{(r)}+\kappa=0$)
%and the positivity of energy density for matters, $\rho\ge0$. 
%Then, we find 
Using Eq.~(\ref{3ricci}) with the maximal slice condition  (i.e., $K=K_{(r)}+\kappa=0$) and assuming 
the positivity of energy density for matters, $\rho\ge0$, then, we find 
\begin{align}
\label{deriv.Geroch2}
\frac{dE}{dy}=\:&\frac{A^{1/2}}{64\pi^{3/2}}\int_{\sigma_y}
\left[2\varphi^{-2}({\cal D} \varphi)^2+16\pi\rho_{\rm tot}+2v_av^a +\frac{3}{2}\kappa^2 \right]dA\ge 0,
\end{align}
where $\rho_{\rm tot}:=\rho+\rho_{\rm gw}$ and $\rho_{\rm gw}$
is defined by 
\begin{equation}
8\pi \rho_{\rm gw}:=\frac{1}{2}(\tilde \kappa_{ab}^2+\tilde k_{ab}^2). 
\end{equation}
The quantity $\rho_{\rm gw}$ may be interpreted as the energy density
of gravitational waves. 
Since the right-hand side of Eq.~\eqref{deriv.Geroch2} is non-negative,
this inequality gives the so-called Geroch monotonicity.
In this paper, we basically follow the argument of Ref.~\cite{wald} for the Penrose inequality. 
The integration of this formula over $y$
in the range $0\le y< \infty$
leads us to one of the key equations of this paper \cite{Lee:2021hft}:
\begin{equation}
m_{\rm ADM}-m_{\rm ext}-\frac{{\cal R}_{A0}}{2}+\frac{{\cal R}_{A0}}{32\pi}\int_{\sigma_0}dAk^2  
\geq \frac{1}{16\pi}\int_0^\infty dy {\cal R}_A(y) \int_{\sigma_y} dAv^av_a,
\label{E-integral}
\end{equation}
where $E(\infty)=m_{\rm ADM}$ is the ADM mass, ${\cal R}_A(y)$ is the areal radius defined by 
${\cal R}_A(y):={\sqrt {A(y)/4\pi}}$ and ${\cal R}_{A0}={\cal R}_A(0)$. 
$m_{\rm ext}$ is a measure of the total rest mass of the matters and 
gravitational waves in the region between $y=0$ and infinity,
\begin{equation} 
m_{\rm ext}:=2\pi \int_0^\infty dy {\cal R}_A(y)^3 \bar \rho_{\rm tot}(y), \label{ext-mass}
\end{equation}
where $\bar \rho_{\rm tot}(y)$ is the surface-averaged total energy density
\begin{equation}
\bar \rho_{\rm tot}(y):=\frac{1}{A}\int_{\sigma_y}dA \rho_{\rm tot}(y). \label{averaged-rho}
\end{equation}

For later discussions,
we define the area-averaged quasilocal angular momentum $\bar J(y)$ by 
\begin{eqnarray}
\Bigl(8\pi \bar J (y)\Bigr)^2:=\frac{A^2}{6\pi} \int_{\sigma_y}v_av^adA \label{ave-j}
\end{eqnarray}
and its minimum value $\bar J_{\rm min}$ in the range $0\le y< \infty$,
\begin{eqnarray}
\bar J_{\rm min}:=\min_{\lbrace \sigma_y \rbrace}\bar J(y). \label{min-j}
\end{eqnarray}
Using Eqs.~\eqref{ave-j} and \eqref{min-j}, the right-hand side of Eq.~\eqref{E-integral} is evaluated as 
\begin{eqnarray}
\frac{1}{16\pi}\int_0^\infty dy {\cal R}_A \int_{\sigma_y} dAv^av_a
& =&
\frac{3}{2}\int_0^\infty dy \frac{\bar J^2}{{\cal R}^3_A}
\nonumber\\
& \geq & \frac{3}{2}{\bar J}_{\rm min}^2 \int_0^\infty \frac{dy}{{\cal R}^3_A}
\ =\
\frac{\bar J^2_{\rm min}}{{\cal R}^3_{A0}},
\label{v^2int}
\end{eqnarray}
where we used the fact that ${\cal R}_{A}={\cal R}_{A0}e^{y/2}$ holds in the IMCF.

\subsection{Vacuum and axisymmetric spacetime}
\label{vacuum}

In the case of vacuum and axisymmetric spacetimes,
further simplification can be done.
Using the axial Killing vector $\phi^a$, one can define 
the Komar angular momentum $J(y)$, 
\begin{equation}
J(y):=\frac{1}{8\pi}\int_{\sigma_y}v^a \phi_a dA. \label{komar-J}
\end{equation}
Note that $J(y)$ does not depend on $y$
when the vacuum Einstein equation holds,
and hence, we simply denote $J$ rather than $J(y)$, hereafter.

The Cauchy-Schwarz inequality and the definition of $J(y)$ give the following inequality,
\begin{equation}
\int_{\sigma_y}v^av_adA  \int_{\sigma_y} \phi^a \phi_a dA \geq \left(\int_{\sigma_y}v^a \phi_a dA \right)^2=(8\pi J)^2. \label{v^2}
\end{equation}
Following Ref.~\cite{Anglada:2017ryp}, let us define
two kinds of radius, 
${\cal R}_\phi (y)$ and ${\cal R}(y)$, by 
\begin{equation}
\frac{8\pi}{3}{\cal R}_\phi^4 :=\int_{\sigma_y}\phi_a \phi^a dA 
\end{equation}
and
\begin{equation}
\frac{1}{{\cal R}^2}:=\frac{3}{2}{\cal R}_{A}\int^\infty_y \frac{{\cal R}_A}{{\cal R}^4_\phi}dy
\end{equation}
respectively. 
Then, Eq.~\eqref{v^2} can be rewritten as  
\begin{equation}
\frac{1}{16\pi}\int_0^\infty dy {\cal R}_A \int_{\sigma_y} dAv^av_a \geq \frac{J^2}{{\cal R}_0^2{\cal R}_{A0}}, \label{intdyRdAvv}
\end{equation}
where ${\cal R}_0:={\cal R}(0)$. 

Note that there is a relation for the radii ${\cal R}_{A}$, ${\cal R}$ and ${\cal R}_\phi$.
In the IMCF, for a convex $\sigma_y$, one can show \cite{Lee:2021hft} 
\begin{equation}
\frac{1}{3} \leq \frac{{\cal R}_{A}^2{\cal R}^2}{{\cal R}_\phi^4} \leq \frac{5}{3}. \label{radius-relation}
\end{equation}
For spherically symmetric spacetimes, it is easy to see that ${\cal R}={\cal R}_A={\cal R}_\phi$ holds. It is also noted
that Eq.~\eqref{v^2} evaluated on $\sigma_0$ can be written as
\begin{equation}
\bar{J}_0^2 \ge \left(\frac{\mathcal{R}_{A0}}{\mathcal{R}_{\phi0}}\right)^4J^2,
\end{equation}
where $\bar{J}_0 := \bar{J}(0)$ and ${\cal R} _{\phi 0} := {\cal R} _\phi (0)$.

\section{Longitudinal attractive gravity probe surface}
\label{lts}

In this section, as generalizations of the loosely trapped surface (LTS) \cite{Shiromizu:2017ego}, we will present two types of 
attractive gravity probe surfaces (AGPSs) as indicators for
the strength of gravity. 

The motivation for introducing the LTS in \cite{Shiromizu:2017ego}
was that the photon sphere or the photon surface can be present only in
highly symmetric spacetimes such as the Schwarzschild spacetime.
Therefore, the generalization of the photon sphere that is
present less symmetric spacetimes was necessary. 
The LTS is defined as follows \cite{Shiromizu:2017ego}:
%====================%
% definition of LTS  %
%====================%
\begin{definition}
In the setup of Sec.~\ref{formula}, an LTS $\sigma_0$ is defined by a compact two-surface satisfying $k>0$ and 
\begin{equation}
{}^{(3)}{\mbox \pounds}_{r}k \ge 0. \label{defLTS} 
\end{equation}
\end{definition}
Let us consider the Schwarzschild spacetime with the metric, 
\begin{equation}
  ds^2=-f(r)dt^2+\frac{dr^2}{f(r)}+r^2\left(d\theta^2+\sin^2\theta d\phi^2\right),
  \label{Schwarzschild-metric}
\end{equation}
where $f(r)=1-2m/r$. For $r$-constant surfaces in $t$-constant hypersurfaces,
the left-hand side of Eq.~\eqref{defLTS} is computed as 
\begin{equation}
{}^{(3)}{\mbox \pounds}_{r}k=-\frac{2}{r^2}\left(1-\frac{3m}{r}\right). \label{1stk-Sch}
\end{equation}
The condition of Eq.~\eqref{defLTS} holds for $r \leq 3m$ and the equality gives us $r=3m$, that is, the location of the 
photon sphere of the Schwarzschild spacetime. 

A generalization of the LTS
was proposed in Ref. \cite{Izumi:2021hlx}
in order to characterize the strength of gravity on surfaces
with an intensity parameter.
This is the original version of 
the attractive gravity probe surface (AGPS), but we will classify it into 
one of the four types of AGPSs in the present paper.  
We will first review the previous studies \cite{Izumi:2021hlx, Lee:2021hft}
on the original version (which we call LAGPS-k in this paper)
in Subsec.~\ref{Sec:LAGPS-k},
and then, present the concept of a new variant, LAGPS-r,
in Sec.~\ref{Sec:LAGPS-r}.

\subsection{LAGPS associated with the mean curvature (LAGPS-k)}
\label{Sec:LAGPS-k}

Inspired by the LTS, we present an attractive gravity probe surface associated with the mean curvature, $k$, as follows \cite{Izumi:2021hlx}:
%=======================%
% definition of LAGPS-k %
%=======================%
\begin{definition}
In the setup of Sec.~\ref{formula}, a longitudinal attractive gravity probe surface associated with 
the mean curvature $k$ (LAGPS-k), denoted by $\sigma_0$, is defined by a compact two-surface satisfying $k>0$ and  
\begin{equation}
{}^{(3)}{\mbox \pounds}_{r}k \ge \alpha k^2, \label{def-lts}
\end{equation}
where $\alpha$ is a constant greater than $-1/2$. 
\end{definition}
Note that we specify the strength of gravity
by the intensity parameter $\alpha$. 
In this definition, the strength of gravity is evaluated 
by comparing the derivative of $k$ along the {\it longitudinal} 
direction to $\sigma_0$ with $k^2$.
Because of this procedure, we call it the {\it longitudinal} AGPS-k (LAGPS-k). 

As an example, let us consider the Schwarzschild spacetime. 
The condition of Eq.~\eqref{def-lts} implies us 
\begin{equation}
r \leq \frac{3+4\alpha}{1+2\alpha}m.
\end{equation}
For $\alpha \to -1/2$, the right-hand side diverges
and the surface with arbitrarily large $r$ can satisfy the condition.
We see that $r \leq 3m$ for $\alpha=0$ and $r \leq 2m$ for 
the $\alpha \to \infty $ limit.
It is easy to check that the prefactor in the right-hand side is a monotonically 
decreasing function of $\alpha$.

In order to discuss the meaning of this definition further,
it is nice to discuss
more general static spherically symmetric spacetimes with the 
metric,
\begin{equation}
ds^2=-f_1(r)dt^2+\frac{dr^2}{f_2(r)}+r^2(d\theta^2+\sin^2 \theta d\phi^2). 
\end{equation}
For this, the condition of Eq.~\eqref{def-lts} becomes 
\begin{equation}
{}^{(3)}{\mbox \pounds}_{r}k -\alpha k^2=\frac{1}{r} \Bigl[f_2'-\frac{2(1+2\alpha)}{r}f_2 \Bigr] \geq 0.
\end{equation}
Supposing the positivity of $f_2$, we rewrite it as
\begin{equation}
f_2' \geq \frac{2(1+2\alpha)}{r}f_2 \geq 0.
\end{equation}
Then, one can see that $f_2$ is a
monotonically increasing function, which corresponds to the attractive
property of gravity, as long as $\alpha>-1/2$ holds. 

%============================%
% Wilmore and PI for LAGPS-k %
%============================%
On the maximal slice $\Sigma$, the surface integral of Eq.~\eqref{liederiv.k3} over the LAGPS-k $\sigma_0$ 
with the condition of Eq.~\eqref{def-lts} implies \cite{Izumi:2021hlx, Lee:2021hft} 
\begin{equation}
\Bigl(1+\frac{4}{3}\alpha \Bigr) \int_{\sigma_0}dAk^2 \leq \frac{16\pi}{3}-\frac{2}{3}\int_{\sigma_0}dA(16\pi \rho_{\rm tot}+2v_av^a). 
\label{willmore-lagpsk}
\end{equation}
From this inequality, we can confirm that $k=0$ on $\sigma_0$ in the limit of $\alpha \to \infty$, that is, $\sigma_0$ is the minimal surface. 
This is consistent with the consequence obtained directly by the condition
of Eq.~\eqref{def-lts}. 

Together with Eqs.~\eqref{E-integral} and \eqref{v^2int}, Eq.~\eqref{willmore-lagpsk} gives us the following 
theorem \cite{Izumi:2021hlx, Lee:2021hft}:
\begin{theorem}
\label{thm.gen.lagpsk}
Let $\Sigma$ be an asymptotically flat spacelike maximal hypersurface equipped by 
the inverse mean curvature flow $\lbrace \sigma_y \rbrace_{y \in {\bf R}}$ with $\sigma_y \approx S^2$, where $\sigma_0$ is an LAGPS-k. 
Assuming that the energy density of matters $\rho$ is non-negative, we have an inequality for the LAGPS-k $\sigma_0$, 
\begin{eqnarray}
m_{\rm ADM}-\Bigl( m_{\rm ext}+\frac{3}{3+4\alpha}m_{\rm int} \Bigr) & \geq & \frac{1+2\alpha}{3+4\alpha}{\cal R}_{A0}
+\frac{1}{{\cal R}^3_{A0}}\Bigl( \bar J_{\rm min}^2+\frac{3}{3+4\alpha} \bar J_0^2 \Bigr) \\
& \geq & \frac{1+2\alpha}{3+4\alpha}{\cal R}_{A0}+2\frac{3+2\alpha}{3+4\alpha} \frac{\bar J_{\rm min}^2}{{\cal R}^3_{A0}},
\label{Penrose-like_inequality_LAGPS-k}
\end{eqnarray}
where 
\begin{eqnarray}
m_{\rm int}:=\frac{4\pi}{3}{\cal R}_{A0}^3 \bar \rho_{{\rm tot}}^{(0)}.
\end{eqnarray}
\end{theorem}
Here, $\rho_{\rm tot}(y)$ is defined in Eq.~\eqref{averaged-rho} 
and $\bar{\rho}_{{\rm tot}}^{(0)}:=\bar{\rho}_{\rm tot}(0)$. 
$m_{\rm int}$ may be regarded as the measure
of the mass of the region enclosed by the surface
estimated from the energy density on the surface. 

%===========================================%
% PI for LAGPS-k in axisymmetric spacetimes %
%===========================================%
For vacuum and axisymmetric spacetimes, using Eq. (\ref{intdyRdAvv}), 
it is possible to refine the Penrose-like inequality
to include the contribution from the angular momentum,
and one has the following theorem \cite{Lee:2021hft}:
\begin{theorem}
\label{thm.axis.1.laspsk}
In vacuum and axisymmetric spacetimes, let $\Sigma$ be an asymptotically flat axisymmetric spacelike maximal 
hypersurface equipped by the inverse mean curvature flow $\lbrace \sigma_y \rbrace_{y \in {\bf R}}$ 
with $\sigma_y \approx S^2$, where $\sigma_0$ is an LAGPS-k.
Assuming that the energy density of matters $\rho$ is non-negative, we have 
an inequality for the LAGPS-k $\sigma_0$,
\begin{eqnarray}
m_{\rm ADM}-\Bigl(m_{\rm gw}^{\rm (ext)}+\frac{3}{3+4\alpha}m_{\rm gw}^{\rm (int)}\Bigr) \geq \frac{1+2\alpha}{3+4\alpha}{\cal R}_{A0}
+\frac{1+\chi_\alpha}{{\cal R}_0^2{\cal R}_{A0}}J^2, \label{pi-axivac-lagpsk}
\end{eqnarray}
where 
\begin{eqnarray}
\chi_\alpha:=\frac{3}{3+4\alpha} \frac{{\cal R}^2_0{\cal R}_{A0}^2}{{\cal R}_{\phi 0}^4},\label{chialpha}
\end{eqnarray}
 and in terms of the surface average of the energy density of
 gravitational waves, 
\begin{eqnarray}
\bar \rho_{{\rm gw}}(y) :=\dfrac{1}{A} \int_{\sigma _y} dA \rho_{{\rm gw}},
\end{eqnarray}
 and $\bar \rho_{{\rm gw}}^{(0)} :=\bar \rho_{{\rm gw}} (0)$,
 $m_{\rm gw}^{\rm (ext)}$ and $m_{\rm gw}^{\rm (int)}$ are defined as
\begin{equation}
m_{\rm gw}^{\rm (ext)}:=2\pi \int_0^\infty dy {\cal R}_A^3 \bar \rho_{\rm gw}, 
\end{equation}
and
\begin{equation}
m_{\rm gw}^{\rm (int)}:=\frac{4\pi}{3}{\cal R}_{A0}^3 \bar \rho_{{\rm gw}}^{(0)}.
\end{equation}
\end{theorem}
Since Eq.~\eqref{radius-relation}
holds for a convex $\sigma_y$ in the IMCF, $\chi_\alpha$ is bounded as \cite{Lee:2021hft}
\begin{equation}
\frac{1}{3+4\alpha} \leq \chi_\alpha \leq \frac{5}{3+4\alpha}. \label{gamma-const}
\end{equation}

\subsection{LAGPS associated with the Ricci scalar (LAGPS-r)}
\label{Sec:LAGPS-r}

Here, we present the variant of the LAGPS-k. 
Let us reconsider Eq. (\ref{1stk-Sch})
in the Schwarzschild case.
One may relate the factor $2/r^2$ in the right-hand side
to the two-dimensional Ricci scalar 
${}^{(2)}R$. Then, inspired by the LTS, it is natural
to define an attractive gravity probe surface associated with the Ricci scalar 
${}^{(2)}R$ as follows:
%=======================%
% definition of LAGPS-r %
%=======================%
\begin{definition}
In the setup of Sec.~\ref{formula}, a longitudinal attractive gravity probe surface associated with the Ricci scalar ${}^{(2)}R$ 
(LAGPS-r), $\sigma_0$, is defined by a two-surface satisfying $k>0$ and 
\begin{equation}
{}^{(3)}{\mbox \pounds}_{r}k\ge-{}^{(2)}R(1-\gamma_L). \label{def-lagps-r}
\end{equation}
\end{definition}
For the Schwarzschild spacetime, the condition of Eq.~\eqref{def-lagps-r} gives us 
\begin{equation}
r \leq 3m/\gamma_L
\end{equation}
for $\gamma_L>0$. Therefore, for $\gamma_L\to +0$, the radius $r$
of the surface can be arbitrarily large, while
for $\gamma_L=1$, the radius must be equal to or smaller than
that of the photon sphere. Although the LAGPS-r with $\gamma_L=3/2$
corresponds to the horizon in the Schwarzschild case,
this statement does not hold in general.
We will come back to this point at the last section.

%============================%
% Wimore funct. for  LAGPS-r %
%============================%
One can show that the Willmore function, $\int_{\sigma_0}k^2dA$, on the LAGPS-r $\sigma_0$ 
is bounded from above. 
From the surface integral of Eq.~\eqref{liederiv.k3} over the LAGPS-r $\sigma_0$ and the condition of Eq.~\eqref{def-lagps-r}, 
we have 
\begin{align}
\int_{\sigma_0}&\left[\left(-\gamma_L+\frac{3}{2}\right){}^{(2)}R-G_{ab}n^an^b+\kappa K_{(r)}\right.\nonumber\\
&\left.+\frac{1}{2}\left(\frac{1}{2}\kappa^2-\tilde{\kappa}_{ab}\tilde{\kappa}^{ab}
-\frac{3}{2}k^2-\tilde{k}_{ab}\tilde{k}^{ab}\right)-v_av^a
-\varphi^{-2}\left({\cal D}\varphi\right)^2
\right]dA\ge 0. \label{k2-integral0-lts}
\end{align}
Assuming the Einstein equation $G_{ab}=8\pi T_{ab}$ to hold for the spacetime $(M,g_{ab})$, $\sigma_0$ to be topologically sphere 
$\sigma_0 \approx S^2$, $k$ to be non-negative $k \geq 0$,
and $\Sigma$ to be maximally sliced, i.e. $K=\kappa+K_{(r)}=0$, 
Eq.~\eqref{k2-integral0-lts} implies 
\begin{align}
\int_{\sigma_0}k^2dA&\le\frac{16\pi}{3}\left(3-2\gamma_L \right)
-\frac{2}{3}\int_{\sigma_0}\left( 16\pi \rho_{{\rm tot}}+2v_av^a+\frac{3}{2}\kappa^2+2\varphi^{-2}\left({\cal D}\varphi\right)^2 \right)dA
\label{wilmoreLAGPSr0} \\
&\le\frac{16\pi}{3}\left( 3-2\gamma_L \right)-\frac{2}{3}\int_{\sigma_0}\left( 16\pi \rho_{{\rm tot}}+2v_av^a \right)dA \label{wilmoreLAGPSr}, 
\end{align} 
where we used the Gauss-Bonnet theorem.
Assuming the non-negativity of the energy density of matters, $\rho\ge 0$,
when $\gamma_L=3/2$, the inequality (\ref{wilmoreLAGPSr0}) tells us 
\begin{equation}
k=\rho_{\rm tot}=v_a=\kappa={\cal D}_a \varphi=0
\end{equation}
on $\sigma_0$. Thus, we have a relatively strong consequence
that $\sigma_0$ is totally geodesic in $\Sigma$. 

%%%%%%%%%%%%%%%%%%%%%%%%%%%%%%%%%%%%%%%%%%%%%%%%%%%%%%%%%%%%%%%%%%%%%%%%%%%%%%%%%%%%%%%%%

%\subsection{Refined inequalities for general case}
%\label{gen.case2}

For the LAGPS-r, Eqs. \eqref{E-integral}, (\ref{v^2int}) and (\ref{wilmoreLAGPSr}) give us the following theorem:
%%%%%%%%
\begin{theorem}
\label{thm.gen.}
Let $\Sigma$ be an asymptotically flat spacelike maximal hypersurface equipped by 
the inverse mean curvature flow $\lbrace \sigma_y \rbrace_{y \in {\bf R}}$ with $\sigma_y \approx S^2$, where $\sigma_0$ is an LAGPS-r. 
Assuming that the energy density of matters $\rho$ is non-negative, we have an inequality for the LAGPS-r $\sigma_0$, 
\begin{align}
m_{\rm ADM}-(m_{\rm int}+m_{\rm ext})
&\geq\frac{\gamma_L}{3}{\cal R}_{A0}+\frac{\bar{J}_0^2+\bar{J}_{\rm min}^2}{{\cal R}_{A0}^3}\nonumber\\
&\geq\frac{\gamma_L}{3}{\cal R}_{A0}+2\frac{\bar{J}_{\rm min}^2}{{\cal R}_{A0}^3}.\label{pi-ineq-LAGPSr}
\end{align}
\end{theorem}
%%%%%%%%

For vacuum and axisymmetric spacetimes,  Eqs. \eqref{E-integral}, (\ref{intdyRdAvv}) and (\ref{wilmoreLAGPSr}) give us the following theorem: 
\begin{theorem}
\label{thm.axis.1}
In vacuum and axisymmetric spacetimes, let $\Sigma$ be an asymptotically flat axisymmetric spacelike maximal 
hypersurface equipped by the inverse mean curvature flow $\lbrace \sigma_y \rbrace_{y \in {\bf R}}$ 
with $\sigma_y \approx S^2$, where $\sigma_0$ is an LAGPS-r. 
%Assuming that the energy density of matters $\rho$ is non-negative, t
Then, we have 
an inequality for the LAGPS-r $\sigma_0$,
\begin{align}
m_{\rm ADM}-(m_{\rm gw}^{\rm (int)}+m_{\rm gw}^{\rm (ext)})
\geq \frac{\gamma_L}{3}{\cal R}_{A0}+\frac{1+\chi_0}{{\cal R}_0^2{\cal R}_{A0}}J^2, \label{pi-ineq2}
\end{align}
where $\chi_0$ is defined by% 
\footnote{This is  equal to the $\alpha=0$ case of $\chi_\alpha$ defined by Eq. (\ref{chialpha}).}
\begin{equation}
\chi_0:=\frac{{\cal R}_0^2{\cal R}_{A0}^2}{{\cal R}_{\phi 0}^4}.
\label{chi0}
\end{equation}
\end{theorem}

\section{Transverse attractive gravity probe surface}
\label{tagps}

In this section,
we discuss the generalizations of the DTTS
proposed in Ref.~\cite{Yoshino:2019dty}
to make it include an intensity parameter for the strength of gravity.
We will present the two types of generalizations,
which we call the transverse attractive gravity 
probe surfaces (TAGPSs). 

%====================%
% definition of DTTS %
%====================%
We begin by reviewing the concept of the DTTS. 
A DTTS was proposed as a generalized concept of the photon sphere or
the photon surface \cite{Claudel:2000yi}.
It is defined as follows \cite{Yoshino:2019dty}:
\begin{definition}
In the setup of Sec.~\ref{formula}, a DTTS $\sigma_0$ is defined by two-surface satisfying the following 
three conditions; 
\begin{align}
&\kappa=0,\label{mom.nonexp}\\
&\mathrm{max}(\bar{K}_{ab}k^ak^b)=0,\label{marg.trans.trap}\\ 
&{}^{(3)}\bar{\mbox \pounds}_{n}\kappa\le 0,\label{accel.cont.}
\end{align}
where $k^a$ is arbitrary null tangent vector to $S$ and the lapse function $N$ is taken to be constant on $\sigma_0$. 
\end{definition}
The conditions of Eqs.~\eqref{mom.nonexp}, \eqref{marg.trans.trap} and \eqref{accel.cont.} are called 
the momentarily non-expanding condition,
the marginally transversely trapping condition, and 
the accelerated contraction condition, respectively
(see more details in \cite{Yoshino:2019dty}).
Here, although we present the definition of the DTTS in the setup of Sec.~\ref{preparation},
the original definition 
in Ref.~\cite{Yoshino:2019dty}
has been given in a broader context:
the timelike hypersurface $S$ needs not be 
orthogonal to the spacelike hypersurface $\Sigma$. 

Here, we focus on the second condition of Eq.~\eqref{marg.trans.trap}
and derive an inequality that is useful in studying the Penrose-like
inequality for the TAGPSs later. 
In general, with $n^a$ and a spacelike unit vector $s^a$ tangent to $\sigma_0$,
the null tangent vector to $S$ can be given by $k^a=n^a+s^a$, and hence,
the left-hand side of Eq. (\ref{marg.trans.trap}) 
is evaluated as  
\begin{eqnarray}
\mathrm{max}(\bar{K}_{ab}k^ak^b) & = &  \mathrm{max}(\bar K_{(n)}+k_{ab}s^as^b-2s^av_a ) \nonumber \\
& = & \bar K_{(n)}+\frac{1}{2}k+ \mathrm{max}(\tilde k_{ab}s^as^b-2s^av_a ) \nonumber \\
& \geq & \bar K_{(n)}+\frac{1}{2}k+ \mathrm{max}(\tilde k_{ab}s^as^b). \label{barKkkineq}
\end{eqnarray}
Here, $n^a$ is fixed and one takes the maximum among various $s^a$ in the $S^1$ directions 
in the second and the third lines. In deriving 
the third line,
we used the fact that for $\pm s^a$ which give
$\mathrm{max}(\tilde k_{ab}s^as^b)$, one of $\mathrm{max}[\tilde k_{ab}s^as^b-2(\pm s^a)v_a]$ must be equal to
or greater than $\mathrm{max}(\tilde k_{ab}s^as^b)$. 
Since $k_{ab}$ is a symmetric tensor, there is the orthogonal basis $\lbrace e_1, e_2 \rbrace$ 
such that $k_{ab}=k_1 (e_1)_a(e_1)_b+k_2 (e_2)_a(e_2)_b$. Then, 
\begin{equation}
\tilde k_{ab}s^as^b=(k_1-k_2)\frac{\cos(2\theta)}{2},
\end{equation}
where $s^a$ is parametrized as $s^a=\cos\theta(e_1)^a + \sin\theta(e_2)^a$.
Thus, we find 
\begin{equation}
\max ( \tilde k_{ab}s^as^b)=\frac{\max (k_1,k_2)-\min (k_1,k_2)}{2} \geq 0,
\end{equation}
and hence,
\begin{eqnarray}
\mathrm{max}(\bar{K}_{ab}k^ak^b) \geq \bar K_{(n)}+\frac{1}{2}k. \label{KkkgeqKn+k}
\end{eqnarray}
Finally, the condition of Eq.~\eqref{marg.trans.trap} implies 
\begin{equation}
-\bar K_{(n)} \geq \frac{1}{2}k.\label{barKn-ineq}
\end{equation}

Below, we extend the concept of the DTTSs
in two ways one by one. The
extended surfaces are called TAGPS-k and TAGPS-r, respectively,
and each of them includes an intensity parameter
for gravity.
In both extensions, 
we keep the first condition of Eq.~\eqref{mom.nonexp}
because it is imposed to fix 
$S$ from $\sigma_0$ (see \cite{Yoshino:2019dty} for detail).
In defining the TAGPS-k, we replace the second condition
and in defining TAGPS-r, we modify the third condition.

\subsection{TAGPS associated with the mean curvature (TAGPS-k)}
\label{Sec:TAGPS-k}

In this subsection, replacing the second condition
of Eq.~\eqref{marg.trans.trap} to a new one, 
we present the concept of a surface as a probe for gravity
strength, which is called the transverse attractive gravity probe
surface associated with the mean curvature (TAGPS-k). 

We first analyze the Schwarzschild spacetime to see how we should modify the condition.  
For simplicity, we adopt $\Sigma$ and $S$ as the 
$t=$~const. and $r=$~const. hypersurfaces, respectively.
In this case, $\kappa=0$ holds everywhere, and
then the conditions of Eqs.~\eqref{mom.nonexp}
and \eqref{accel.cont.} are trivially satisfied, while 
the condition of Eq.~\eqref{marg.trans.trap} becomes non-trivial. 
Adopting $k^a=n^a+s^a$, where $s^a$ is a spatial unit vector tangent to $S$ and orthogonal to $\Sigma$, one can compute 
$\bar K_{ab} k^a k^b$ as 
\begin{equation}
\bar K_{ab}k^a k^b=\bar K_{(n)}+\frac{1}{2}k=\frac{1}{r{\sqrt {f}}}\left(1-\frac{3m}{r}\right),\label{schbarKkk}
\end{equation}
where we used $\bar K_{(n)}=-(1/2)f^{-1/2}\partial_r f$ and $k=2f^{1/2}/r$. 
Therefore, the condition of Eq.~\eqref{marg.trans.trap} implies
$r=3m$, that is, the location of the photon 
sphere of the Schwarzschild spacetime. 

As a trial for the extension of the DTTS to include an intensity parameter
for gravity,
instead of the condition of Eq.~\eqref{marg.trans.trap}, we may impose 
\begin{equation}
\mathrm{max}(\bar K_{ab}k^a k^b) \leq -\beta k \label{trialdefTAGPSk}
\end{equation}
and $\beta$ is a constant larger than $-1/2$. Using Eq. (\ref{KkkgeqKn+k}), Eq. (\ref{trialdefTAGPSk}) leads to
\begin{equation}
\bar K_{(n)} \leq -\frac{2\beta+1}{2}k. \label{barKnleqk}
\end{equation}
%If one takes the limit of $\beta \to -1/2$, the above becomes $\bar K_{(n)}=-mf^{-1/2}r^{-2} \leq 0$. 
%This is satisfied when supposes the positivity of $m$ reflecting to the attractiveness of gravity. 
For the Schwarzschild spacetime, Eq. (\ref{barKnleqk}) gives us  
\begin{equation}
r \le \frac{3+4\beta}{1+2\beta} m.
\end{equation}
It is easy to see that the coefficient of the right-hand side is monotonically decreasing function of $\beta$. 
The right-hand side becomes $2m$  for the limit of $\beta\to\infty$ and $3m$ for the $\beta=0$ case. 
Since we can regard $\beta$ in Eq.~\eqref{trialdefTAGPSk} as an intensity parameter for gravity, 
we will adopt the condition of Eq.~\eqref{trialdefTAGPSk} instead of
Eq.~\eqref{marg.trans.trap} for the TAGPS-k. 
Note that the null tangent $k^a$ must be normalized as the above argument for the Schwarzschild spacetime.  

%=======================%
% definition of TAGPS-k %
%=======================%
\begin{definition}
In the setup of Sec.~\ref{formula}, a TAGPS-k $\sigma_0$ is defined by a compact two-surface satisfying the 
three conditions; 
\begin{align}
&\kappa=0,\label{mom.nonexp3}\\
&\mathrm{max}(\bar{K}_{ab}k^ak^b)\leq -\beta k,\label{defDTAGPS}\\ 
&{}^{(3)}\bar{\mbox \pounds}_{n}\kappa\le 0,\label{accel.cont.3}
\end{align}
where $k^a$ is arbitrary null tangent vector to $S$ such that $p_{ab}k^b=n_a$ holds, 
and the lapse function $N$ is taken to be constant on $\sigma_0$. 
\end{definition}

%=======================%
%    Wilmore function   %
%=======================%
Let us investigate the properties of the TAGPS-k. 
We will show that the Willmore function, $\int_{\sigma_0}k^2dA$, on the TAGPS-k $\sigma_0$ 
is bounded from above.
To show this, we consider the surface integral of Eq.~\eqref{liederiv.kappa3} over $\sigma_0$ 
and then,
using the inequality of Eq.~\eqref{barKnleqk},
which originates from 
the condition of Eq.~\eqref{defDTAGPS}, we have 
\begin{equation}
\int_{\sigma_0}\left[-\frac{1}{2}{}^{(2)}R-G_{ab}r^ar^b+\left( \beta+\frac{3}{4}\right)k^2
-\frac{1}{2}\left(\tilde k_{ab} \tilde k^{ab}+\tilde{\kappa}_{ab}\tilde{\kappa}^{ab}\right)+v_av^a\right]dA\le 0, \label{k2-integral0}
\end{equation}
where we used the fact that the time lapse function $N$ is constant% 
\footnote{
Inequality \eqref{k2-integral0} is achieved even if $N$ is not constant, 
because the contribution involving $N$ is positive, 
\begin{eqnarray}
\int_{\sigma_0} N^{-1} {\cal D}^2 N \, dA 
=\int_{\sigma_0} \left( N^{-1} {\cal D} N \right)^2 \, dA \ge 0. 
\end{eqnarray}
Therefore, Theorems \ref{thm.gen.-dtagps} and \ref{thm.axis.-dtagps} hold even for  non-constant $N$.
} 
on $\sigma_0$. 
Note that on $\sigma_0$, $\kappa_{ab}$ is a traceless quantity because of the condition of Eq.~\eqref{mom.nonexp3}, i.e. 
$\kappa_{ab}=\tilde{\kappa}_{ab}$. 
We now assume the Einstein equation $G_{ab}=8\pi T_{ab}$ to hold for the spacetime $(M,g_{ab})$, $\sigma_0$ to be topologically sphere 
$\sigma_0 \approx S^2$, and $k$ to be nonnegative $k \geq 0$ (we will soon adopt the inverse mean curvature flow where we have to impose $k \geq 0$). 
Then, Eq.~\eqref{k2-integral0} implies 
\begin{equation}
\Bigl(1+\frac{4}{3}\beta \Bigr)\int_{\sigma_0}k^2dA \leq \frac{16\pi}{3}
+\frac{2}{3}\int_{\sigma_0}(16\pi P_{r}^{\rm (tot)}-2v_av^a)dA. \label{wilmoreTAGPSk}
\end{equation}
In the above, $P_{r}^{\rm (tot)}$ is the total radial pressure, $P_{r}^{\rm (tot)}=P_r+P_{r}^{\rm (gw)}$, where $P_r$ is the 
radial pressure for matters defined by $P_r:=T_{ab}r^ar^b$ and $P_{r}^{\rm (gw)}$ is the radial pressure of the gravitational wave, 
\begin{equation}
8\pi P_{r}^{\rm (gw)}=\frac{1}{2}(\tilde k_{ab} \tilde k^{ab}+\tilde \kappa_{ab}\tilde \kappa^{ab})=8\pi \rho_{\rm gw}.
\end{equation}
In the limit of $\beta \to \infty$, we can see that $k=0$ holds on $\sigma_0$, that is, $\sigma_0$ is the minimal surface. 
This consequence is directly expected from the condition of
Eq.~\eqref{defDTAGPS}. 

For the TAGPS-k, Eqs. \eqref{E-integral}, \eqref{v^2int} and \eqref{wilmoreTAGPSk} give us the following theorem:
%%%%%%%%
\begin{theorem}
\label{thm.gen.-dtagps}
Let $\Sigma$ be an asymptotically flat spacelike maximal hypersurface equipped by 
the inverse mean curvature flow $\lbrace \sigma_y \rbrace_{y \in {\bf R}}$ with $\sigma_y \approx S^2$, where $\sigma_0$ is a TAGPS-k. 
Assuming that the energy density of matters $\rho$ is non-negative, we have an inequality for the TAGPS-k $\sigma_0$, 
\begin{align}
m_{\rm ADM}+\frac{3}{3+4\beta}{p}_{r}^{\rm (int)}-m_{\rm ext}&\geq\frac{1+2\beta}{3+4\beta}{\cal R}_{A0}
+\frac{1}{{\cal R}_{A0}^3}\left(\frac{3}{3+4\beta}\bar{J}_0^2+\bar{J}_{\rm min}^2\right)\nonumber\\
&\ge\frac{1+2\beta}{3+4\beta}{\cal R}_{A0}+\frac{2(3+2\beta)}{3+4\beta}\frac{\bar{J}_{\rm min}^2}{{\cal R}_{A0}^3},\label{pi-ineq-dtagps}
\end{align}
where 
\begin{eqnarray}
{p}_{r}^{\rm (int)}:=\frac{4\pi}{3}{\cal R}_{A0}^3 \bar{P}_{r0}^{\rm (tot)} \label{p-int-dtagps}
\end{eqnarray}
and $\bar P_{r0}^{\rm (tot)}$ is the area-averaged total pressure evaluated on $\sigma_0$
\begin{eqnarray}
\bar P_{r0}^{\rm (tot)}=\frac{1}{A_0}\int_{\sigma_0}P_{r}^{\rm (tot)}dA.
\end{eqnarray}
\end{theorem}
%%%%%%%%%%

In vacuum and axisymmetric cases, we obtain the following areal inequality for the TAGPS-k.
\begin{theorem}
\label{thm.axis.-dtagps}
In vacuum and axisymmetric spacetimes, let $\Sigma$ be an asymptotically flat axisymmetric spacelike maximal 
hypersurface equipped by the inverse mean curvature flow $\lbrace \sigma_y \rbrace_{y \in {\bf R}}$ 
with $\sigma_y \approx S^2$, where $\sigma_0$ is a TAGPS-k. 
%Assuming that the energy density of matters $\rho$ is non-negative, 
Then, we have 
an inequality for the TAGPS-k $\sigma_0$,
\begin{align}
m_{\rm ADM}+\frac{3}{3+4\beta}{p}_{\rm gw}^{\rm (int)}-m_{\rm gw}^{\rm (ext)}
\geq \frac{1+2\beta}{3+4\beta}{\cal R}_{A0}+\frac{1+\chi_\beta}{{\cal R}_0^2{\cal R}_{A0}}J^2,\label{pi-ineq2-dtagps}
\end{align}
where 
\begin{equation}
\chi_\beta:=\frac{3}{3+4\beta}\frac{{\cal R}_0^2{\cal R}_{A0}^2}{{\cal R}_{\phi 0}^4}
\label{gammaalpha}
\end{equation}
and
\begin{equation}
{p}_{\rm gw}^{\rm (int)}:=\frac{4\pi}{3}{\cal R}_{A0}^3 \bar P_{r0}^{\rm (gw)}. 
\end{equation}
\end{theorem}
Since $\chi_\beta$ has the same expression as $\chi_\alpha$ introduced in Eq. (\ref{pi-axivac-lagpsk}), 
the same constraint as Eq. (\ref{gamma-const}) holds
but $\alpha$ being replaced by $\beta$.  

%There are two types for the shpae of $\sigma_y$ with the principal curvatures $\lambda_\phi$ and $\lambda_\theta$.
%For the oblate case, $\lambda_\theta \geq \lambda_\phi >0$,  
%\begin{equation}
%{\cal R}_\phi \geq {\cal R}_A
%\end{equation}
%or for the prolate case, $ \lambda_\phi \geq \lambda_\theta >0 $, 
%\begin{equation}
%{\cal R}_\phi \leq {\cal R}_A.
%\end{equation}

\subsection{TAGPS associated with the Ricci scalar (TAGPS-r)}
\label{dtts}

In this subsection, we give a different extension of the DTTS
to include an intensity parameter for gravity.
In contrast to the TAGPS-k, we keep the first and the second conditions,
and examine the modification
to third condition of Eq.~\eqref{accel.cont.}. 
For this purpose, we first examine a photon surface $S$ of the Schwarzschild spacetime that consists of a collection of
worldlines of transversely emitted photons from an $r$-constant sphere in a $t$-cosntant hypersurface,  
where the conditions of Eqs.~\eqref{mom.nonexp} and \eqref{marg.trans.trap} are satisfied (see more 
details in \cite{Yoshino:2019dty}).
The Lie derivative of $\kappa$ with respect to $n^a$ is% 
\footnote{This quantity can be evaluated independently of the choice
  of $\Sigma$ as long as $\Sigma$ is spherically symmetric.}
\begin{equation}
\label{barliederivkappa}
{}^{(3)}\bar{\mbox \pounds}_{n}\kappa=\frac{2}{r^2}\left(1-\frac{3m}{r}\right).
\end{equation}
Therefore, the condition of Eq.~\eqref{accel.cont.} implies  $r \le 3m$. 
The equality of Eq. (\ref{accel.cont.}) states $r=3m$, which is exactly
the same as the location of the photon sphere of the 
Schwarzschild spacetime. 

As a trial of extension of the DTTS to include an intensity parameter for
gravity strength, instead of the condition of Eq.~\eqref{accel.cont.},
we may impose 
\begin{equation}
{}^{(3)}\bar{\mbox \pounds}_{n}\kappa\le{}^{(2)}R(1-\gamma_T), \label{trialdefTAGPSr}
\end{equation}
where $\gamma_T$ is a constant. For the Schwarzschild spacetime,
this condition is reduced to 
\begin{equation}
r \le \frac{3m}{\gamma_T}.
\end{equation}
Therefore, in the limit $\gamma_T\to +0$,
$r$ can be arbitrarily large, and 
$r \leq 3m$ for the $\gamma_T=1$ case,
and $r\leq 2m$ for the $\gamma_T=3/2$ case. 
The parameter $\gamma_T$ in Eq.~\eqref{trialdefTAGPSr}
is regarded as the intensity parameter for gravity.
For this reason, 
we will adopt the condition of Eq.~\eqref{trialdefTAGPSr} instead of
Eq.~\eqref{accel.cont.} for the
new definition of the TAGPS, which we call the
transverse AGPS associated with the Ricci scalar
(TAGPS-r).

%=======================%
% definition of TAGPS-r %
%=======================%
\begin{definition}
In the setup of Sec.~\ref{formula}, a TAGPS-r $\sigma_0$ is defined by a compact two-surface satisfying the 
three conditions; 
\begin{align}
&\kappa=0,\label{mom.nonexp2}\\
&\mathrm{max}(\bar{K}_{ab}k^ak^b)=0,\label{marg.trans.trap2}\\ 
&{}^{(3)}\bar{\mbox \pounds}_{n}\kappa\le{}^{(2)}R(1-\gamma_T),\label{defTAGPSr}
%&\int_{\sigma_0}dA{}^{(3)}\bar{\mbox \pounds}_{n}\kappa\le8\pi(1-\gamma),\label{defgDTTS}
\end{align}
where $k^a$ is arbitrary null tangent vector to $S$ such that $p_{ab}k^b=n_a$ holds, 
and the lapse function $N$ is taken to be constant on $\sigma_0$. 
\end{definition}

Here, several remarks are added.
In the above definition, the quantity $k$ (the trace of the extrinsic
curvature of $\sigma_0$ in the spacelike hypersurface $\Sigma$)
is not used. In this sense, the concept of the TAGPS-r is free from the choice
of the spacelike hypersurface. Physically,
this reflects the fact that the TAGPS-r is defined
only in terms of the behavior of transversely emitted photons
from $\sigma_0$. 
For this reason, similarly to the DTTS, the definition of
the TAGPS-r needs not be restricted to the setup of Sec.~\ref{preparation}.
In particular, the TAGPS-r has the coordinate invariance 
in the following sense:
If $\sigma_0$ is obtained as the TAGPS-r in the spacelike hypersurface
$\Sigma$, on a different spacelike hypersurface
$\Sigma^\prime$ which crosses $\Sigma$ exactly at $\sigma_0$,
we can obtain $\sigma_0$ as the TAGPS-r as well.

%==========================%
% Wimore func. for TAGPS-r %
%==========================%
We now discuss the areal inequality for the TAGPS-r.
The analysis below relies on the setup of Sec.~\ref{preparation}. 
We show that the Willmore function, $\int_{\sigma_0}k^2dA$,
on the TAGPS-r $\sigma_0$ 
has an upper bound. Considering the surface integral of Eq.~\eqref{liederiv.kappa3} over $\sigma_0$, 
and then, imposing the condition of Eq.~\eqref{defTAGPSr}, we have 
\begin{equation}
\int_{\sigma_0}\left[ \left( \gamma_T-\frac{3}{2}\right){}^{(2)}R-G_{ab}r^ar^b-k\bar{K}_{(n)}+\frac{1}{2}\left(k^2-k_{ab}k^{ab}
-\tilde{\kappa}_{ab}\tilde{\kappa}^{ab}\right)+v_av^a\right]dA\le 0. \label{k2-integral3}
\end{equation}
In the above, we used
the fact that the time lapse function $N$ is constant%
\footnote{As with the case of inequality \eqref{k2-integral0}, inequality \eqref{k2-integral3} can be obtained even if $N$ is not constant. 
Therefore, Theorems \ref{thm.gen.tagps-r} and \ref{thm.axis.} hold even for non-constant $N$.
} 
on $\sigma_0$. 
Note that on $\sigma_0$, $\kappa_{ab}$ is a traceless quantity because of the condition of Eq.~\eqref{mom.nonexp2}, i.e. 
$\kappa_{ab}=\tilde{\kappa}_{ab}$. 
We now assume the Einstein equation $G_{ab}=8\pi T_{ab}$ to hold for the spacetime $(M,g_{ab})$,
$\sigma_0$ to be topologically sphere 
$\sigma_0 \approx S^2$, and $k$ to be non-negative  $k \geq 0$
(we will soon adopt the inverse mean curvature flow where we have to 
impose $k \geq 0$). Then, with Eq.~\eqref{barKn-ineq}, Eq.~\eqref{k2-integral3} implies 
\begin{eqnarray}
\int_{\sigma_0}k^2dA \leq \frac{16\pi}{3}(3-2\gamma_T)+\frac{2}{3}\int_{\sigma_0}(16\pi P_{r}^{\rm (tot)}-2v_av^a)dA . \label{wilmoreTAGPSr}
\end{eqnarray}
If $P_{r}^{\rm (tot)} \leq 0$ is assumed on $\sigma_0$, 
\begin{equation}
k=P_{r}^{\rm (tot)}=v_a=0
\end{equation}
holds for $\gamma_T=3/2$. 

%\subsubsection{Refined inequalities for general case}
%\label{gen.case}

For the TAGPS-r, Eqs. \eqref{E-integral}, (\ref{v^2int}) and \eqref{wilmoreTAGPSr} give us the following theorem:
\begin{theorem}
\label{thm.gen.tagps-r}
Let $\Sigma$ be an asymptotically flat spacelike maximal hypersurface equipped by 
the inverse mean curvature flow $\lbrace \sigma_y \rbrace_{y \in {\bf R}}$ with $\sigma_y \approx S^2$, where $\sigma_0$ is the TAGPS-r. 
Assuming that the energy density of matters $\rho$ is non-negative, we have the following inequality for the TAGPS-r $\sigma_0$:
\begin{align}
m_{\rm ADM}+{p}_{r}^{\rm (int)}-m_{\rm ext}&\geq\frac{\gamma_T}{3}{\cal R}_{A0}
+\frac{\bar{J}_0^2+\bar{J}_{\rm min}^2}{{\cal R}_{A0}^3}\nonumber\\
&\ge\frac{\gamma_T}{3}{\cal R}_{A0}+2\frac{\bar{J}_{\rm min}^2}{{\cal R}_{A0}^3}\label{pi-ineq}
\end{align}
\end{theorem}
%\begin{proof}
%Equation~\eqref{wilmoreTAGPSr} is rewritten as 
%\begin{equation}
%\int_{\sigma_0}k^2dA \leq \frac{16\pi}{3}\left( 3-2\gamma_T \right)-\frac{32\pi}{3}A_0 \bar P_{r,{\rm tot}0}
%-32\pi \frac{\bar{J}^2}{{\cal R}_{A0}^4},\label{k-ineq2}
%\end{equation}
%where we used $\bar J$ defined by Eq.~\eqref{ave-j}. 
%Then, applying Eqs.~\eqref{k-ineq2} and \eqref{v^2int} to Eq.~\eqref{E-integral}, we obtain Eq.~\eqref{pi-ineq}. 
%\end{proof}

%%%%%%%%%%%%%%%%%%%%%%%%%%%%%%%%%%%%%%%%%%%%%%%%%%%%%%%%%%%%%%%%%%%%%%%%%%%%%%%%%%%
%\subsection{vacuum and axially symmetric cases}
%\label{vac.axis.case}

For vacuum and axisymmetric spacetimes, Eqs.~\eqref{E-integral}, (\ref{intdyRdAvv}) and \eqref{wilmoreTAGPSr} give us the following theorem: 
\begin{theorem}
\label{thm.axis.}
In vacuum and axisymmetric spacetimes, let $\Sigma$ be an asymptotically flat axisymmetric spacelike maximal 
hypersurface equipped by the inverse mean curvature flow $\lbrace \sigma_y \rbrace_{y \in {\bf R}}$ 
with $\sigma_y \approx S^2$, where $\sigma_0$ is the TAGPS-r. 
Then, we have 
the inequality for the TAGPS-r $\sigma_0$,
\begin{align}
m_{\rm ADM}+p_{\rm gw}^{\rm (int)}-m_{\rm gw}^{\rm (ext)}
\geq \frac{\gamma_T}{3}{\cal R}_{A0}+\frac{1+\chi_0}{{\cal R}_0^2{\cal R}_{A0}}J^2.\label{pi-ineq3}
\end{align}
\end{theorem}
%\begin{proof}

%\end{proof}

\section{summary and discussion}
\label{summary}

In this paper, we have reexamined the attractive gravity probe surface (AGPS) proposed in Ref.~\cite{Izumi:2021hlx}, and 
then proposed the four types of AGPSs including the original one, that is, the longitudinal AGPS associated with the mean 
curvature (LAGPS-k) and the Ricci scalar (LAGPS-r), and the transversely AGPS
associated with the mean curvature (TAGPS-k) and 
the Ricci scalar (TAGPS-r). For these four AGPSs, under certain conditions, we have shown the Penrose-like inequalities that include the 
contribution from the angular momentum and matters. The differences between
them can be found in several aspects, as summarized below.

The first difference is that
the masslike quantity $m_{\rm int}$ appears in the Penrose-like inequalities
for the LAGPSs, Eqs.~\eqref{Penrose-like_inequality_LAGPS-k}
and \eqref{pi-ineq-LAGPSr}, while the pressurelike quantity $p_{\rm int}$
appears in the Penrose-like inequalities
for the TAGPSs, Eqs.~\eqref{pi-ineq2-dtagps}
and \eqref{pi-ineq}. This reflects the fact that
the LAGPSs are defined in terms of the geometrical
quantities in the spatial direction, while the TAGPSs
are defined in terms of the geometrical
quantities in the timelike direction.
The contribution from the pressure of the inside region, if it is positive, 
tends to relax the Penrose-like inequalities for the two types of the TAGPSs,
while the contribution from 
the energy density of the inside makes those inequalities stronger
for the two types of the LAGPSs.

Next, we focus on the case of a minimal surface in vacuum and axisymmetric spacetimes.
For an LAGPS-k with $\alpha \to \infty$  and a TAGPS-k with $\beta \to \infty$ that correspond to the minimal surface, on the one hand, we have the same inequality
\begin{align}
m_{\rm ADM}-m_{\rm gw}^{\rm (ext)} \geq \frac{{\cal R}_{A0}}{2}+\frac{J^2}{{\cal R}_0^2{\cal R}_{A0}} \label{minimal1}
\end{align}
from Eqs.~\eqref{pi-axivac-lagpsk} and \eqref{pi-ineq3}. 
This reproduces Anglada's result \cite{Anglada:2017ryp}. 
Moreover, the square of Eq.~\eqref{minimal1} gives 
\begin{align}
\left(m_{\rm ADM}-m_{\rm gw}^{\rm (ext)}\right)^2 
\geq \Bigl( \frac{{\cal R}_{A0}}{2} \Bigr)^2+\frac{J^2}{{\cal R}_0^2}+\Bigl( \frac{J^2}{{\cal R}_0^2{\cal R}_{A0}}\Bigr)^2 
\geq \Bigl( \frac{{\cal R}_{A0}}{2} \Bigr)^2+\frac{J^2}{{\cal R}_0^2}.
\end{align}
Then, we have obtained the inequality similar to that of Eq.~\eqref{conj2}. 
On the other hand, 
for an LAGPS-r and a TAGPS-r, setting $\gamma_L=\gamma_T=3/2$ which corresponds to the minimal surface in the case of the Schwarzschild spacetime, we have 
\begin{align}
m_{\rm ADM}-m_{\rm gw}^{\rm (ext)} \geq \frac{{\cal R}_{A0}}{2},
\end{align}
where we used the fact that the angular momentum and $\rho_{\rm gw}=P_{r}^{\rm (gw)}$ vanishes on the LAGPS-r and the TAGPS-r  
when $\gamma_L=\gamma_T=3/2$. Thus, we cannot refine the inequalities
by including the contribution from the angular 
momentum. Furthermore, as argued later, the surfaces
with $\gamma_L=\gamma_T=3/2$ do not 
correspond to the minimal surface in general. Thus, the LAGPS-r and the TAGPS-r
do not characterize the minimal surface in a direct way.

Note that the application of the arithmetic-geometric mean for our inequalities gives us new inequalities. 
For example, for the LAGPS-k, Eq. (\ref{pi-axivac-lagpsk}) gives \cite{Lee:2021hft}
\begin{align}
m_{\rm ADM} \geq 2F_\alpha \frac{|J|}{{\cal R}_0},
\end{align}
where
\begin{equation}
  F_\alpha=\left[\frac{(1+\chi_\alpha)(1+2\alpha)}{(3+4\alpha)}\right]^{1/2}.
\end{equation}
We also point out that, applying the arithmetic-geometric mean
after taking the square for the inequality of Eq.~\eqref{pi-axivac-lagpsk}, 
we have 
\begin{eqnarray}
m_{\rm ADM}^2 & \geq & \left( \frac{1+2\alpha}{3+4\alpha}{\cal R}_{A0}\right)^2+2(1+\chi_\alpha)\frac{1+2\alpha}{3+4\alpha} \frac{J^2}{{\cal R}_0^2}
+\left(\frac{1+\chi_\alpha}{{\cal R}_0^2{\cal R}_{A0}}J^2 \right)^2 \nonumber \\
 & \geq & \left( \frac{1+2\alpha}{3+4\alpha}{\cal R}_{A0}\right)^2+2(1+\chi_\alpha)\frac{1+2\alpha}{3+4\alpha} \frac{J^2}{{\cal R}_0^2} \nonumber \\
 & \geq & \eta_\alpha \frac{{\cal R}_{A0}}{{\cal R}_0}|J|, \label{m-J-ineq}
\end{eqnarray}
where the second line was derived in Ref. \cite{Lee:2021hft} and 
\begin{eqnarray}
\eta_\alpha:=\left[\frac{2(1+2\alpha)}{3+4\alpha}  \right]^{3/2}(1+\chi_\alpha)^{1/2}.
\end{eqnarray}
For the prolate LAGPS-k, we can show 
\begin{eqnarray}
\left( \frac{{\cal R}_{A0}}{{\cal R}_0} \right)^2  
& = & \frac{3}{2}{\cal R}_{A0}^3\int_0^\infty \frac{{\cal R}_A^4}{{\cal R}_\phi^4} \frac{1}{{\cal R}_A^3}dy \nonumber \\
& \geq &  \frac{3}{2}{\cal R}_{A0}^3\int_0^\infty \frac{1}{{\cal R}_A^3}dy \nonumber \\
& = & 1,
\end{eqnarray}
where we used the fact that ${\cal R}_A /{\cal R}_\phi \geq 1$ holds for the prolate LAGPS-k\footnote{We can show ${\cal R}_{A0}/{\cal R}_0 \leq 1$ for the oblate LAGPS-k.} (see Ref.~\cite{Lee:2021hft}) in the second line and 
${\cal R}_{A}={\cal R}_{A0}e^{y/2}$ in the third line. Then, Eq. (\ref{m-J-ineq}) gives us 
\begin{eqnarray}
m_{\rm ADM}\geq  \eta_\alpha^{1/2} |J|^{1/2}. \label{m-J-ineq2}
\end{eqnarray}
For the limit $\alpha \to \infty$ where
the LAGPS-k approaches 
the minimal surface, this inequality reduces to $m_{\rm ADM}\geq |J|^{1/2}$.
This inequality has been shown in Refs.~\cite{Dain2006, Dain2008} for spacetimes close to the extreme Kerr solution in a different way.
We would like to stress that Eq. (\ref{m-J-ineq2}) works for a wide class
of spacetimes.

%===========<FIGURE1>============%
%
\begin{figure}[tb]
\centering
\includegraphics[width=0.45\textwidth,bb=0 0 293 194]{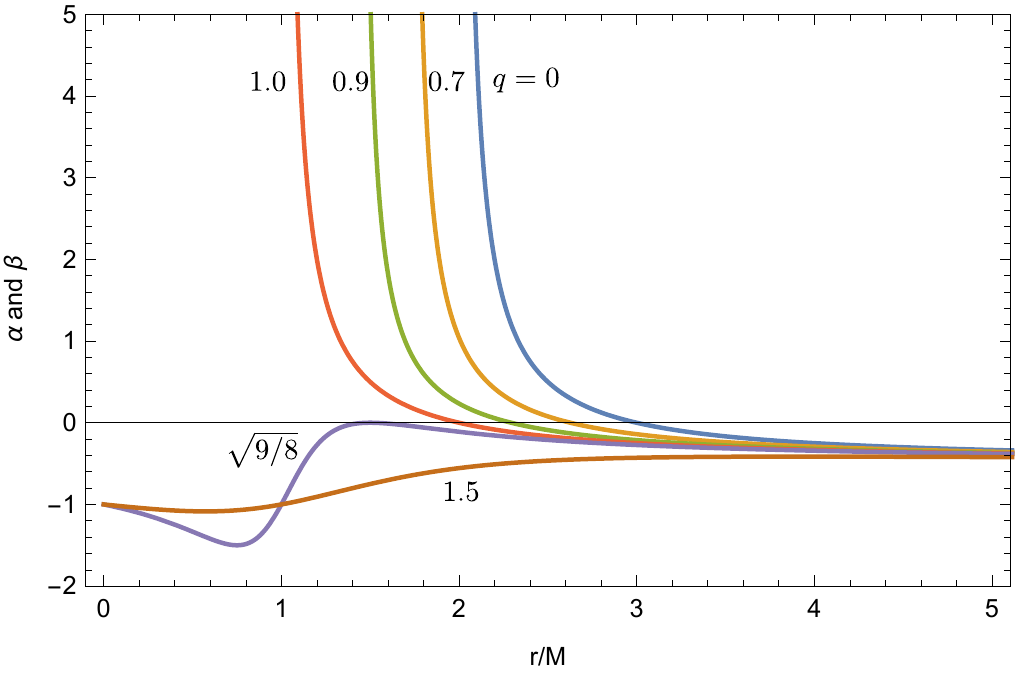}
\includegraphics[width=0.45\textwidth,bb=0 0 293 194]{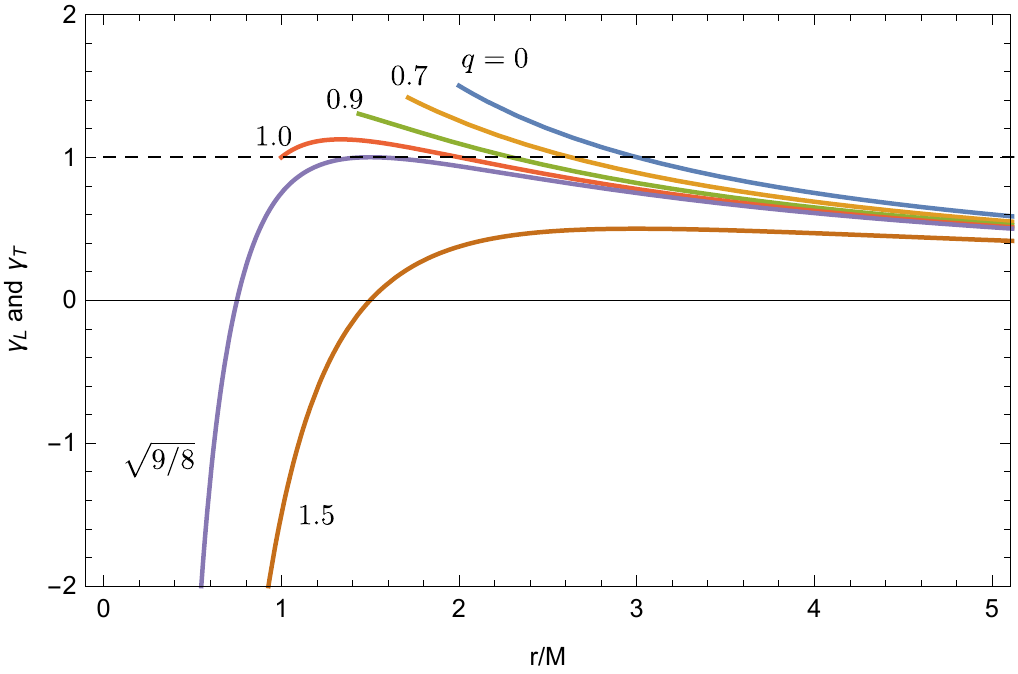}
\caption{Behavior of the intensity parameters
  of the four AGPSs in a Reissner-Nordstr\"om spacetime.
  The cases $q=0$, $0.7$, $0.9$, $1.0$, $\sqrt{9/8}$, and $1.5$
  are shown. Left panel: The values of $\alpha$ and $\beta$
  for the marginal LAGPS-k and the marginal TAGPS-k
  as functions of the radius $r$. Here, $\alpha=\beta$ holds.
  Right panel: The behavior of $\gamma_L$ and $\gamma_T$
  for the marginal LAGPS-k and the marginal TAGPS-k.
  Here, $\gamma_L=\gamma_T$ holds.
}
\label{AGPS-edited}
\end{figure}
%
%=================================%

As seen above, the LAGPS-r and the TAGPS-r could not include
the effect of the angular momentum in the Penrose-like inequality
for $\gamma_L=\gamma_T=3/2$. We do not consider that these definitions
fail because of this. 
Rather, the ways of characterizing the intensity of gravity
are different. 
We would like to discuss this point by
adopting a Reissner-Nordstr\"om spacetime
as an example. The metric of the Reissner-Nordstr\"om spacetime
is given by Eq.~\eqref{Schwarzschild-metric} with
\begin{equation}
f(r)=1-\frac{2m}{r}+\frac{Q^2}{r^2},
\end{equation}
and it is convenient to introduce the nondimensional parameter
$q=Q/m$ to specify the solution. For a fixed parameter $q$,
we adopt $\Sigma$ to be the $t$-constant slice.
Then, 
on each $r$-constant surface, 
we can evaluate the values of $\alpha$, $\gamma_L$, $\beta$, 
and $\gamma_T$ such that the conditions for the four AGPSs
are marginally satisfied (i.e., the cases where the equalities hold).
Figure~\ref{AGPS-edited} shows the behavior
of the intensity parameters for selected values of $q$.
The curves of $\alpha$ and $\beta$ are the same (the left panel),
and the curves of $\gamma_L$ and $\gamma_T$ are the same (the right panel).
For $0\le q\le 1$, each curve has the left edge which corresponds
to the location of the horizon.
On the one hand,
$\alpha$ and $\beta$ are zero at the photon sphere,
and diverge on the horizon, as expected. On the other hand,
although $\gamma_L$ and $\gamma_T$ are one at the photon sphere,
they do not have a constant value on the horizon.

These results are interpreted as follows. Since
the quantity $k$ is included directly in the definitions
of the LAGPS-k and the TAGPS-k,
it can represent both of the photon sphere and the minimal surface
with specific values of $\alpha$ and $\beta$.
However, the quantity $k$ is not included
in the definitions of the LAGPS-r and the TAGPS-r.
Let us take the extremal case $q=1$ for simplicity.
It is well known that the extremal black hole has
an infinitely long throat, and hence the quantity
${}^{(3)}\mbox \pounds_rk$ that appears in the definition of the LAGPS-r,
Eq.~\eqref{def-lagps-r}, is obviously zero.
Therefore $\gamma_L=1$ holds in this case,
and this is the consequence of reflecting the geometric
structure in the radial direction.
As for the TAGPS-r, 
the spacetime possesses two photon spheres
for $1<q<\sqrt{9/8}$ and the inner stable photon sphere
is reduced to the horizon generator in the limit $q\to 1$.
Therefore, the value of ${}^{(3)}\bar{\mbox \pounds}_{n}\kappa$
that appears in the definition
of the TAGPS-r, Eq.~\eqref{defTAGPSr},
is zero, and hence, $\gamma_T=1$. 
This is also the consequence of the behavior of transversely emitted photons.
This discussion indicates that the four AGPSs characterize 
the gravity intensity in different ways.
Although the intensity parameters $\gamma_L$ and $\gamma_T$
do not have a specific value on the horizon,
it may be related to the surface gravity $\kappa$ 
which also vanishes on the horizon of the extremal black hole.

The four concepts of the AGPSs introduced in this paper
would have different meanings, and hence,
would be useful in different contexts.
Clarification of the meaning of each concept further, in particular by exploring 
spacetimes with rotating black holes, would be our interesting
remaining problem.

\acknowledgments

T. S. and K. I.  are supported by Grant-Aid for Scientific Research from Ministry of Education, 
Science, Sports and Culture of Japan (Nos. 17H01091, JP21H05182). T. S., K. I. and H. Y are 
supported by JSPS(JP21H05189). T. S. is also supported by JSPS Grants-in-Aid for Scientific Research (C) (JP21K03551). 
K.~I. is also supported by JSPS Grants-in-Aid for Scientific Research (B) (JP20H01902).
H. Y. is in part supported by JSPS KAKENHI Grant Numbers JP22H01220,
and is partly supported by Osaka Central Advanced Mathematical Institute 
(MEXT Joint Usage/Research Center on Mathematics and Theoretical Physics JPMXP0619217849).

\appendix

\section{Unification of LAGPSs}

In Secs.~\ref{Sec:LAGPS-k} and \ref{Sec:LAGPS-r} of the main text,
we proposed the LAGPS-k and the LAGPS-r, separately. 
In this appendix, we briefly describe the unified treatment of
these two LAGPSs. The setup and assumptions are 
same as Sec.~\ref{preparation}.
The conditions (\ref{def-lts}) and (\ref{def-lagps-r})
in the definitions of the LAGPS-k and the LAGPS-r are unified as 
\begin{align}
{}^{(3)}{\mbox \pounds}_r k \geq \alpha k^2-{}^{(2)}R\left(1-\gamma_L \right)\label{heAGPS}.
\end{align}
Then, the inequality for the Willmore function becomes 
\begin{equation}
\left(1+\frac{4}{3}\alpha\right)\int_{\sigma _0}k^2dA\le\frac{16\pi}{3}\left(3-2\gamma_L \right)
-\frac{2}{3}\int_{\sigma _0}\left( 16\pi \rho_{{\rm tot}}+2v_av^a \right)dA.
\end{equation}
As a result, we obtain the Penrose-like inequality that includes
two intensity parameters, $\alpha$ and $\gamma_L$,
\begin{align}
m_{\rm ADM}-\left(\frac{3}{3+4\alpha}{m}_{\rm int}+m_{\rm ext}\right)\ge\frac{\gamma_L
+2\alpha}{3+4\alpha}{\cal R}_{A0}+\frac{1}{{\cal R}_{A0}^3}\left(\frac{3}{3+4\alpha}\bar{J}_0^2+\bar{J}_{\rm min}^2\right).
\end{align}

\section{Unification of TAGPSs}

In Secs.~\ref{Sec:TAGPS-k} and \ref{dtts} of the main text,
we proposed the TAGPS-k and the TAGPS-r, separately. 
In this appendix, we briefly describe the unified treatment of
these two TAGPSs. The setup and assumptions are 
same as Sec.~\ref{preparation}.
As for the momentarily non-expanding condition, Eqs.~\eqref{mom.nonexp3}
and \eqref{mom.nonexp2} are the same.
As the marginally transversely trapping condition, we adopt the
one for the TAGPS-k, Eq.~\eqref{defDTAGPS},
and as the accelerated contraction condition,
we adopt the one for the TAGPS-r, Eq.~\eqref{defTAGPSr},
in order to include both parameters.
Summarizing these conditions, we have 
\begin{align}
&\kappa=0,\label{mom.nonexp-he}\\
&\mathrm{max}(\bar{K}_{ab}k^ak^b)\leq -\beta k,\label{marg.trans.trap-he}\\ 
&{}^{(3)}\bar{\mbox \pounds}_{n}\kappa\le {}^{(2)}R\left(1-\gamma_T \right).\label{accel.cont.-he}
\end{align}
Then, the inequality for the Willmore function is derived as 
\begin{equation}
\left(1+\frac{4}{3}\beta\right)\int_{\sigma_0}k^2dA\le\frac{16\pi}{3}\left(3-2\gamma_T \right)
+\frac{2}{3}\int_{\sigma_0}\left( 16\pi P_{r,{\rm tot}}-2v_av^a \right)dA.
\end{equation}
As a result, we obtain the Penrose-like inequality that includes
two intensity parameters, $\beta$ and $\gamma_T$, 
\begin{align}
m_{\rm ADM}+\frac{3}{3+4\beta}{p}_{r}^{{\rm (int)}}-m_{\rm ext}\ge\frac{\gamma_T+2\beta}{3+4\beta}{\cal R}_{A0}+\frac{1}{{\cal R}_{A0}^3}\left(\frac{3}{3+4\beta}\bar{J}_0^2+\bar{J}_{\rm min}^2\right).
\end{align}

%%%%%%%%%%%%%%%%%%%%%%%%%%%%%%%%%%%%%%%%%%%%%%%%%%%%%%%%%%%%%%%%%%%%%%%%%%%%%%%%%%%%%%%%%

%%%%%%%%%%%%%%%%%%%%%%%%%%%%%%%%%%%%%%%%%%%%%%%%%%%%%%%%%%%%%%%%%%%%%%%%%%%%%%%%%%%%%%%%%
    %---------   References   ---------%
%\newpage

\end{document}